\documentclass
[showkeys,showpacs,preprintnumbers,nofootinbib]
{revtex4} 
\usepackage{graphicx}
\usepackage{bm}
\begin{document}   
\preprint{CYCU-HEP-09-16}
\title{Studies on the polarization transfer coefficients $C_{x}$ and $C_{z}$
\\in the $\vec{\gamma} p\to K^{+}\vec{\Lambda}(1520,3/2^{-})$ reaction  process}
\author{Seung-il Nam}
\email{sinam@cycu.edu.tw}
\affiliation{Department of Physics, Chung-Yuan Christian University (CYCU), 
Chung-Li 32023, Taiwan} 
\date{\today}
\begin{abstract}  
We investigate the polarization transfer coefficients $C_{(x,z)}$ for the $\vec{\gamma} p\to K^{+}\vec{\Lambda}(1520,3/2^{-})$ reaction process, in which the photon is polarized circularly and the $\Lambda(1520,3/2^{-})$ along the $x$- or $z$-axis. To this end, we employ the effective Lagrangian method at tree level and the gauge-invariant form factor scheme. In addition to the Born terms, $(s,u,t_{K},t_{K^{*}})$-channels and contact term, we include the nucleon resonance $D_{13}(2080)$ in the $s$-channel. We compute the $C_{(x,z)}$ as functions of $\theta_{K}$  as well as $E_{\mathrm{cm}}$. It turns out that the $K^{*}$-exchange and $D_{13}$ contributions are negligible within available experimental and theoretical inputs for them. In contrast, we observe that the contact and $K$-exchange contributions play dominant roles for determining the $C_{(x,z)}$. Especially, the $K$-exchange enhances the transverse polarization transfer $C_{x}$ considerably.
\end{abstract} 
\pacs{13.75.Cs, 14.20.-c}
\keywords{polarization transfer coefficient, $\Lambda(1520)$-photoproduction, circularly polarized photon}  
\maketitle
\section{Introduction}
Photoproduction of a meson off the nucleon target is one of the useful methods to study strongly interacting systems in terms of the hadronic degrees of freedom. For instance, it enables us to investigate the interaction structure of hadrons, baryon resonances, and various meaningful physical quantities for hadronic properties with the clear probe, photon. Especially, since the photon can be polarized linearly or circularly, various combinations of the polarizations of the photon, target nucleon, and recoil baryon can be identified as polarization observables. Among the possible polarization observables, we are interested in the polarization transfer coefficients $C_{(x,z)}$ using the circularly polarized photon in the present work. These physical quantities measure the spin projections of the recoil baryon on the photon-beam axis ($z$-axis) or the other axis on the reaction plane, ($x$-axis). In other words, they indicate how much the initial helicity is transferred to the recoil baryon, which is polarized in a certain direction. One may refer to Refs.~\cite{Fasano:1992es,Anisovich:2007bq} for details on the polarization observables for the pseudoscalar-meson photoprodcution with the spin-$1/2$ recoil baryon in the final state.

We note that a series of experiments to measure the $C_{(x,z)}$ for the $\gamma p\to K^{+}\Lambda(1116,1/2^{+})$ and $K^{+}\Sigma^{0}(1192,1/2^{+})$ reaction processes was already carried out by the CLAS collaboration~\cite{McNabb:2003nf,Bradford:2005pt,Bradford:2006ba}, in which the photon beam is almost $100\%$ circularly polarized, via the Bremsstrrahlung from the polarized electron beam. It was reported that the $C_{z}$ turns out to be almost unity, whereas the $C_{x}$ zero, as functions of the outgoing kaon angle in the center of mass frame. A possible explanation on this strong polarization transfer in the  longitudinal direction was provided in terms of the $\phi$-meson and vector-meson dominance (VMD), considering a strangeness-quark pair ($\bar{s}s$) inside the $\phi$-meson~\cite{Schumacher:2008xw}. In Ref.~\cite{Anisovich:2007bq}, the isobar partial wave analysis was applied to these data. From the experiments, it is also found that various theoretical-hadronic models for the $C_{(x,z)}$ were still not good enough to reproduce the data, although a simple Regge model showed poor but qualitative agreement~\cite{Bradford:2006ba}. One of the reasons for these unsatisfying theoretical results is that the groud-state hyperon-photoproudcton is dominated much by the nucleon- and hyperon-resonance contributions, resulting in complicated interferences between the relevant contributions~\cite{Janssen:2001pe}. A Born approximation plus the quark-level scattering amplitudes was done in Ref.~\cite{Henley:2009qw}, showing qualitative agreement to a certain extend. 

In contrast to a great amount of the experimental and theoretical studies for the ground-state hyperon-photoproductions, only a few have been done for the spin-$3/2$ hyperons, especially for  $\Lambda(1520,3/2^{-})\equiv\Lambda^{*}$~\cite{Barber:1980zv,Nam:2005uq,Nam:2006cx,Toki:2007ab,Muramatsu:2009zp,Kohri:2009xe}. Interestingly, in our previous work, it turned out that the $\Lambda^{*}$-photoproduction is not affected much from the nucleon resonances~\cite{Nam:2005uq}. In other words, only the Born terms reproduce the presently available experimental data qualitatively well~\cite{Barber:1980zv,Muramatsu:2009zp}. Moreover, it turned out that the contact interaction of $\gamma N$-$K\Lambda^{*}$ (see Fig.~\ref{Feyn}) dominates the reaction process, resulting in a large difference in the cross sections off the neutron and proton targets. Hence, taking into account the energetic progresses on the polarization experiments in experimental facilities such as the CLAS and LEPS, and less theoretical studies for the polarization observables for the $\Lambda^{*}$-photoproduction, in the present work, we are motivated to study the polarization transfer coefficients $C_{(x,z)}$ for the $\gamma p\to K^{+}\Lambda^{*}$ reaction process. To this end, we employ the  effective Lagrangian method and gauge-invariant form factor scheme as done in Refs.~\cite{Nam:2005uq,Nam:2006cx}. Here, we consider the baryon-pole ($s$- and $u$-channels) and the pseudoscalar ($K$) and vector ($K^{*}$) kaon exchanges in the $t$-channel, in addition to the contact term, which is essential to preserve the gauge invariance of the scattering amplitude, i.e., the Ward-Takahashi (WT) identity. Although the present reaction process is thought as an almost resonance-free process, we include a nucleon resonance $D_{13}(2080,3/2^{-})$ among possible resonances, which can couple to the $K\Lambda^{*}$ state, as suggested in the quark-model calculations~\cite{Capstick:1998uh,Capstick:2000qj}. 

We compute the invariant amplitude in the center of mass (cm) frame with the form factors, which are included in such a way to maintain the WT identity. All the calculations are performed firstly in the $(x',y',z')$-coordinate, in which the outgoing kaon is aligned along the $z'$-axis and the $x'$-$z'$ plane is taken as the reaction plane. Then, by virtue of the three-vector nature of the polarization transfer coefficients, we perform a finite rotation from the $(x',y',z')$-coordinate to the $(x,y,z)$-coordinate, in which the $z$-axis is chosen to be along the incident photon beam as in the CLAS experiment. Since the recoil baryon we are interested in possesses spin-$3/2$, the polarization transfer coefficients can be defined by each spin state as $C_{(x,z),1/2}$ and $C_{(x,z),3/2}$, being different from the usual spin-$1/2$ recoil baryon. According to the helicity conservation, the $C_{x,(1/2,3/2)}$ and $C_{z,(1/2,3/2)}$ become zero and unity in the collinear limit, in which the recoil-baryon polarization is parallel to to incident photon beam. 

It turns out that the $K^{*}$-exchange contribution appears to be very small, if we assume that $|g_{K^{*}N\Lambda^{*}}|\approx g_{KN\Lambda^{*}}\approx11$. Similarly, the nucleon-resonance $D_{13}(2080)$ also gives only negligible contributions to the $C_{(x,z)}$ as long as we use presently available theoretical and experimental information for it. We observe that most dominant contributions come from the contact term and the $K$-exchange in the $t$-channel. We note that this contact term and $K$-exchange dominance is a consequence of our gauge-invariant form factor scheme, where the $s$- and $u$-channels are suppressed more than others. Especially, the $K$-exchange enhances the transverse transfer $C_{x}$, whereas the contact term produces oscillating curves for the $C_{(x,z)}$ around zero and unity. There appears a structure in $C_{(x,z)}$ in the vicinity of $\cos\theta_{K}\approx0.5$ due to the $K$-exchange. As for the energy dependence, the $|C_{z}|$ decreases with respect to the center of mass energy $E_{\mathrm{cm}}$, whereas the $|C_{x}|$ increases mainly due to the $K$-exchange contribution. As a result, the $C_{(x,z)}$ varies much from zero and unity, being different from those observed from the ground-state hyperon, $\Lambda(1116)$ and $\Sigma(1192)$, photoproductions experimentally. 

We organize the present work as follows: In Section II, we provide the formalism to compute the $\Lambda^{*}$-photoproduciton. We make a brief review on the unpolarized observables computed with the present effective Lagrangian method in Section III. In Section IV, we define the polarization transfer coefficients and show a simple example to compute the $C_{(x,z)}$. The numerical results for the $C_{(x,z),(1/2,3/2)}$ as functions of $\cos\theta_{K}$ and $E_{\mathrm{cm}}$ are given in Section V. Finally, Section VI is devoted to summary and conclusion. 

\section{Formalism}
\begin{figure}[t]
\includegraphics[width=10cm]{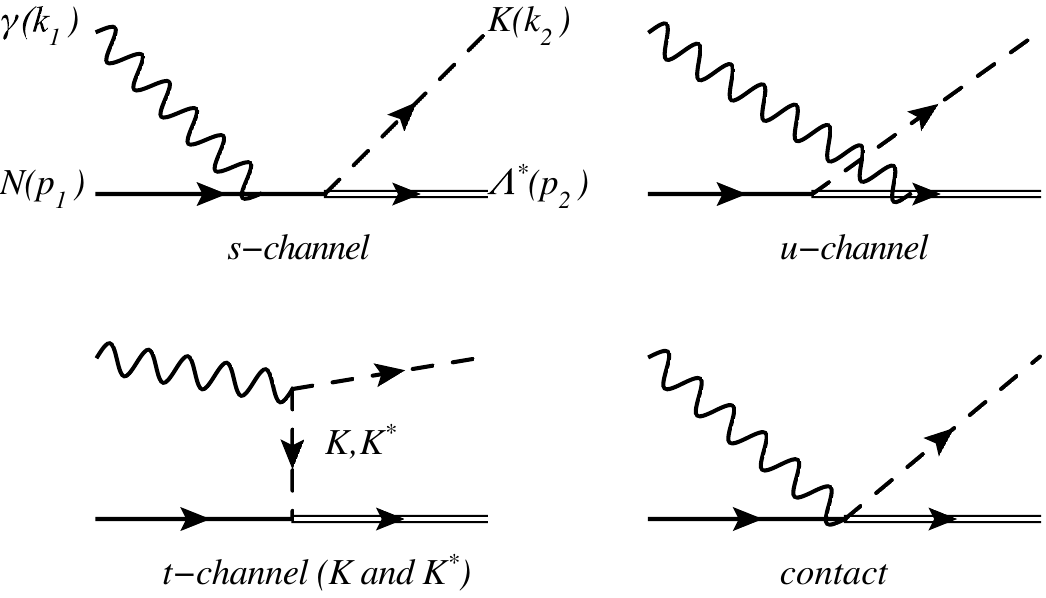}
\caption{Relevant Feynman diagrams for the $\Lambda(1520)$-photoproduction off the nucleon target.}       
\label{Feyn}
\end{figure}

In this section, we provide the basic formalism for computing the $\gamma N\to K\Lambda^{*}$ reaction process. The relevant Feynman diagrams at tree level are shown in Fig.~\ref{Feyn}, in which we define the four momenta of the particles involved. Note that one needs to consider the contact term to conserve the Ward-Takahashi (WT) identity in the present treatment of the spin-$3/2$ particle. In our previous work, it turned out that it dominates the reaction process~\cite{Nam:2005uq}. 

The relevant electromagnetic (EM) and strong interactions are defined as follows:
\begin{eqnarray}
\label{eq:GROUND}
{\cal L}_{\gamma KK}&=&
ie_K\left[(\partial^{\mu}K^{\dagger})K-(\partial^{\mu}K)K^{\dagger}
\right]A_{\mu}+{\rm h.c.},
\nonumber\\
{\cal L}_{\gamma NN}&=&-\bar{N}\left[e_N\rlap{/}{A}-\frac{e_Q\kappa_N}{4M_N}\sigma\cdot F\right]N,
\nonumber\\
\mathcal{L}_{\gamma \Lambda^*\Lambda^*}&=&
\frac{e_Q\kappa_{\Lambda^*}}{4M_{\Lambda^*}}\bar{\Lambda}^{*\mu}
(\sigma\cdot F)\Lambda^*_{\mu}+{\rm h.c.},
\nonumber\\
\mathcal{L}_{KN\Lambda^*}&=&\frac{g_{KN\Lambda^*}}{M_{\Lambda^*}}
\bar{\Lambda}^{*\mu}\partial_{\mu}K\gamma_5N\,+{\rm h.c.},
\nonumber\\
{\cal L}_{\gamma
KN\Lambda^*}&=&-\frac{ie_Ng_{KN\Lambda^*}}{M_{\Lambda^*}}
\bar{\Lambda}^{*\mu}
A_{\mu}K\gamma_5N+{\rm h.c.},
\end{eqnarray}
where the $e_{h}$ and $e_{Q}$ stand for the electric charge of the hadron $h$ and unit charge, respectively. The $A$, $K$, $K^{*}$, $N$, and $\Lambda^{*}$ are the fields for the photon, kaon, vector kaon, nucleon, and $\Lambda^{*}$. As for the spin-$3/2$ fermion field, we make use of the Rarita-Schwinger (RS) vector-spinor field~\cite{Rarita:1941mf,Nath:1971wp} (see Appendix for details). We use the notation $\sigma\cdot F=\sigma_{\mu\nu}F^{\mu\nu}$, where $\sigma_{\mu\nu}=i(\gamma_{\mu}\gamma_{\nu}-\gamma_{\nu}\gamma_{\mu})/2$ and the EM field strength tensor $F^{\mu\nu}=\partial^{\mu}A^{\nu}-\partial^{\nu}A^{\mu}$. The $\kappa_{N,\Lambda^{*}}$ denote the anomalous magnetic momentum for the nucleon and the $\Lambda^{*}$. Although the spin-$3/2$ $\Lambda^{*}$ has four different electromagnetic form factors~\cite{Gourdin:1963ub}, we only take into account the dipole one, ignoring the quadrupole and octopole ones, since their contributions are small. From the experimental data, one can easily find that $g_{\gamma K^{*\pm}K^{\mp}}=0.254/\mathrm{GeV}$~\cite{Amsler:2008zzb}. The strength of the $g_{KN\Lambda^{*}}$ can be calculated using the full and partial decay widths: $\Gamma_{\Lambda^{*}}\approx15.6$ MeV and $\Gamma_{\Lambda^{*}\to\bar{K}N}/\Gamma_{\Lambda^{*}}\approx0.45$. The decay amplitude for $\Lambda^{*}\to \bar{K}N$ reads:
\begin{equation}
\label{eq:decay}
\Gamma_{\Lambda^{*}\to\bar{K}N}=\frac{g^{2}_{KN\Lambda^*}
|{\bm p}_{\bar{K}N}|}
{4\pi M^{2}_{\Lambda^{*}}M^{2}_{K}}
\left(\frac{1}{4}\sum_{\mathrm{spin}}|
\mathcal{M}_{\Lambda^{*}\to\bar{K}N}|^{2}\right)
,\,\,\,\,\,i\mathcal{M}_{\Lambda^{*}\to\bar{K}N}  
=\bar{u}(q_{\Lambda^{*}})\gamma_{5}q_{\bar{K}}^{\mu}u_{\mu}(q_{N}),
\end{equation}
where the ${\bm p}_{\bar{K}N}$ indicates the three momentum of the the decaying particle which can be obtained by the K\"allen function for a decay $1\to2,3$~\cite{Amsler:2008zzb}:
\begin{equation}
\label{eq:kollen}
{\bm p}_{23}=\frac{\sqrt{[M^{2}_{1}-(M_{2}+M_{3})^{2}][M^{2}_{1}-(M_{2}-M_{3})^{2}]}}{2M_{1}}.
\end{equation}
Here, the $M_{i}$ stands for the mass of the $i$-th particle. Substituting the experimental information into Eq.~(\ref{eq:decay}) and using Eq.~(\ref{eq:kollen}), one is lead to $g_{KN\Lambda^*}\approx11$. The scattering amplitudes for the present reaction process can be evaluated straightforwardly using the interactions in Eq.~(\ref{eq:GROUND}):
\begin{eqnarray}
\label{eq:AMP}
i\mathcal{M}_{s}&=&-\frac{g_{KN\Lambda^*}}{M_{K}}
\bar{u}^{\mu}_2k_{2\mu}{\gamma}_{5} 
\left[\frac{e_{N}[(\rlap{/}{p}_{1}+M_{p})F_{c}+\rlap{/}{k}_{1}F_{s}]}
{(k_{1}+p_{1})^{2}-M^{2}_{p}}\rlap{/}{\epsilon}
-\frac{e_{Q}\kappa_{p}}{2M_{p}}
\frac{(\rlap{/}{k}_{1}+\rlap{/}{p}_{1}+M_{p})F_{s}}
{(k_{1}+p_{1})^{2}-M^{2}_{p}}
\rlap{/}{\epsilon}\rlap{/}{k}_{1}\right]u_1,
\cr
i\mathcal{M}_{u}&=&-\frac{e_{Q}g_{KN\lambda}\kappa_{\Lambda^*}F_{u}}
{2M_{K}M_{\Lambda}}
\bar{u}^{\mu}_{2}\rlap{/}{k}_{1}\rlap{/}{\epsilon}
\left[\frac{(\rlap{/}{p}_{2}-\rlap{/}{k}_{1}-M_{\Lambda^{*}})}
{(p_{2}-k_{1})^{2}-M^{2}_{\Lambda^{*}}}
 \right]
k_{2\mu}\gamma_{5}u_{1},
\cr
i\mathcal{M}^{K}_{t}&=&\frac{2e_{K}g_{KN\Lambda^*}F_{c}}{M_K}
\bar{u}^{\mu}_2
\left[\frac{(k_{1\mu}-k_{2\mu})(k_{2}\cdot\epsilon)}
{(k_{1}-k_{2})^{2}-M^{2}_{K}} \right]
\gamma_{5}u_1,
\cr
i\mathcal{M}^{K^{*}}_{t}&=&
-\frac{ig_{\gamma{K}K^*}g_{K^{*}NB}F_{v}}{M_{K^{*}}}   
\bar{u}^{\mu}_{2}\gamma_{\nu}
\left[\frac{(k^{\mu}_{1}-k^{\mu}_{2})g^{\nu\sigma}-
(k^{\nu}_{1}-k^{\nu}_{2})g^{\mu\sigma}
}{(k_{1}-k_{2})^{2}-M^{2}_{K^*}}\right]
(\epsilon_{\rho\eta\xi\sigma}k^{\rho}_{1}
\epsilon^{\eta}k^{\xi}_{2})u_1,
\cr
i\mathcal{M}_{\mathrm{cont.}}
&=&\frac{e_{K}g_{KN\Lambda^*}F_{c}}{M_K}
\bar{u}^{\mu}_2\epsilon_{\mu}{\gamma}_{5}u_1,
\label{amplitudes}  
\end{eqnarray}
where $\epsilon$, $u_{1}$, and ${u}^{\mu}_{2}$ denote the photon polarization vector, nucleon spinor, and RS vector-spinor. Since we are interested in the polarization transfer coefficients, the photon is circularly polarized. For definiteness, we write it in the right- and left-handed ones:
\begin{equation}
\label{eq:pol}
\epsilon^{\mathrm{right}}_{\mu}=-\frac{1}{\sqrt{2}}(0,1,i,0),
\,\,\,\,
\epsilon^{\mathrm{left}}_{\mu}=\frac{1}{\sqrt{2}}(0,1,-i,0).
\end{equation}
Here, we assume that the photon is polarized circularly $100\%$ for simplicity. In further discussions, we will consider only the right-handed polarization hereafter. 

Since the hadrons are not point-like particles, it is necessary to introduce form factors, representing their spatial distributions. It is rather technical to include the form factors conserving the WT identity. For this purpose, we employ the form factor scheme developed in Refs.~\cite{Haberzettl:1998eq,Davidson:2001rk}. This scheme preserves not only the Lorentz invariance but also the crossing symmetry of the amplitude, on top of the WT identity. Moreover, it satisfies the on-shell condition for the form factors: $F(q^{2}=0)=1$. In this scheme, the form factors $F_{s,t,u}$ are defined generically as:
\begin{equation}
\label{eq:form}
F_{s,t,u}=\frac{\Lambda^{4}}{\Lambda^{4}+(s-M^{2}_{s,t,u})^{2}}.
\end{equation}
Here, the subscripts $(s,t,u)$ indicate the Mandelstam variables, whereas the $M_{s,t,u}$ the masses of the off-shell particles in the $(s,t,u)$-channels. The $\Lambda$ stands for a phenomenological cutoff parameter, which will be determined by matching with experimental data. The {\it common} form factor $F_{c}$, which plays a critical role to keep the WT identity, reads: 
\begin{equation}
\label{eq:fc}
F_{c}=F_{s}+F_{t}-F_{s}F_{t}.
\end{equation}
It is clear that the $F_{c}$ satisfies the on-chell condition for arbitrary values of $s$ and $t$. 

Now, we are in a position to consider the nucleon-resonance contributions. To date, we have not had much experimental information on the nucleon resonances, which couple to the $\Lambda^{*}$. The situation seems much worse for the hyperons decaying into $\gamma\Lambda^{*}$. Only some theoretical calculations have provided information on the decays~\cite{Capstick:1998uh,Capstick:2000qj}. Being different from the ground state $\Lambda(1116)$-photoproduction, in which nucleon and hyperon resonances play important roles to produce the data~\cite{Janssen:2001pe}, it is interesting that only the Born terms looks still good to explain the presently available experimental data for the $\Lambda^{*}$-photorpdocution~\cite{Barber:1980zv,Muramatsu:2009zp}, although more dedicated experiments may reveal more complicated structures from unknown sources. Keeping this situation in mind, we attempt to include nucleon-resonance contribution using theoretical information from the relativistic constituent-quark model calculations~\cite{Capstick:1998uh}. Among the possible nucleon resonances given in Ref.~\cite{Capstick:1998uh}, we only choose the $D_{13}(2080)$ with the two-star confirmation ($**$). Although the $S_{11}(2080)$ and $D_{15}(2300)$ may contribute to the present reaction process among the possible resonances given in Ref.~\cite{Capstick:1998uh}, we drop them, since the $S_{11}$ is still in poor confirmation and the $D_{15}$ may increase theoretical uncertainties in dealing with the spin-$5/2$ Lorentz structure~\cite{Choi:2007gy}. In the theoretical calculations of Ref.~\cite{Capstick:1998uh}, $N^{*}(1945,3/2^{-})$ is identified as the $D_{13}(2080)$. However, we will use the available experimental values for the $D_{13}(2080)$, not the theoretical ones, expect for the theoretical value for the partial decay width $\Gamma_{D_{13}\to K\Lambda^{*}}\approx6.76$ GeV.  In order for computing the resonance contribution, we first define the transition and strong interactions for it as follows: 
\begin{eqnarray}
\label{eq:lreso}
\mathcal{L}_{\gamma NN^{*}}&=&
-\frac{ie_{Q}\mu_{\gamma NN^{*}}}{M_{N^{*}}}
\bar{N}^{*}_{\mu} \gamma_{\nu}F^{\mu\nu}N+\mathrm{h.c.},
\cr
\mathcal{L}_{KN^{*}\Lambda^{*}}&=&
\frac{g_{KN^{*}\Lambda^{*}}}{M_{K}}
\bar{\Lambda}^{*}_{\mu}\gamma_{5}\rlap{/}{\partial}
KN^{*\mu}+\mathrm{h.c.},
\end{eqnarray}
where the $N^{*}$ denotes the field for the $D_{13}(2080)$. The magnetic coupling constant $e_{Q}\mu_{\gamma NN^{*}}$ can be computed using the following relation:
\begin{equation}
\label{eq:eee}
|e_{Q}\mu_{\gamma NN^{*}}|=
\left[\frac{6M_{N}M^{2}_{N^{*}}}{(3M^{2}_{N^{*}}+M^{2}_{N})} 
\right]^{\frac{1}{2}}
\left[(A^{N^{*}}_{1/2})^{2}+(A^{N^{*}}_{3/2})^{2}\right]^{\frac{1}{2}},
\end{equation}
Here, the $A^{N^{*}}_{1/2,3/2}$ stand for the helicity amplitudes given experimentally as~\cite{Amsler:2008zzb}: 
\begin{equation}
\label{eq:hel}
A^{p^{*}}_{1/2}=(-0.020\pm0.008)/\sqrt{\mathrm{GeV}},\,\,\,\, 
A^{p^{*}}_{3/2}=(-0.017\pm0.011)/\sqrt{\mathrm{GeV}}.
\end{equation}
Combining Eqs.~(\ref{eq:eee}) and (\ref{eq:hel}), one obtains $|e_{Q}\mu_{\gamma NN^{*}}|\approx0.035$. Although one can determine the sign of the $e_{Q}\mu_{\gamma NN^{*}}$~\cite{Choi:2007gy}, we will use it as a free parameter, which plays the role of a relative phase factor. Employing the theoretical value for the $\Gamma_{D_{13}\to K\Lambda^{*}}$ and using the following relation, one can obtain $g_{KN^{*}\Lambda^{*}}\approx30.14$:
\begin{equation}
\label{eq:gamma}
\Gamma_{N^{*}\to K\Lambda^{*}}=
\frac{g^{2}_{KD_{13}\Lambda^{*}}|{\bm p}_{K\Lambda^{*}}|}
{4\pi M_{D_{13}}M^{2}_{K}}
\left[2E_{K}(E_{\Lambda^{*}}E_{K}+{\bm p}^{2}_{K\Lambda^{*}})
-M^{2}_{K}(E_{\Lambda^{*}}+M_{\Lambda^{*}})\right]
\approx6.76\,\mathrm{GeV}.
\end{equation}
Considering all the ingredients discussed so far, the scattering amplitude for the $D_{13}$ in the $s$-channel can be written as follows:
\begin{equation}
\label{eq:re}
i\mathcal{M}^{*}_{s}=\mathcal{C}
\frac{e_{Q}\mu_{\gamma ND_{13}}\,g_{KD_{13}\Lambda^{*}}F_{s}}
{M_{K}M_{D_{13}}}\bar{u}^{\mu}_{2}\gamma_{5}\rlap{/}{k}_{2}
\left[\frac{(\rlap{/}{k}_{1}+\rlap{/}{p}_{1}+M_{N^{*}})
(k_{1\mu}\rlap{/}{\epsilon}-\rlap{/}{k}_{1}\epsilon_{\mu})}
{(k_{1}+p_{1})^{2}-M^{2}_{D_{13}}-iM_{N^{*}}
\Gamma_{D_{13}}} \right]u_{1},
\end{equation}
where the $\mathcal{C}$ denotes a free parameter for the relative phase factor as mentioned above. We will take $\mathcal{C}=\pm1$. The $\Gamma_{D_{13}}$ is the full decay width and has large experimental uncertainty, $\Gamma_{D_{13}}=(87\sim1075)$ MeV~\cite{Amsler:2008zzb}. As a trial, we choose $\Gamma_{D_{13}}\approx500$ MeV for the numerical calculations, although there is no firm theoretical ground for this choice at this moment. We verify that there are only small differences observed even for sizable change in the $\Gamma_{D_{13}}$. If the $\Gamma_{D_{13}}$ becomes far narrower as lower than $\sim100$ MeV, the resonance contribution may become obvious. However, we consider that such a narrow nucleon resonance seems unlikely, unless there are unusual production mechanisms for the resonance like the exotics. Thus, we confined ourselves to the condition that $\Gamma_{D_{13}}\gtrsim300$ MeV as usual nucleon resonances in the region behave. 

\begin{figure}[t]
\includegraphics[width=7.0cm]{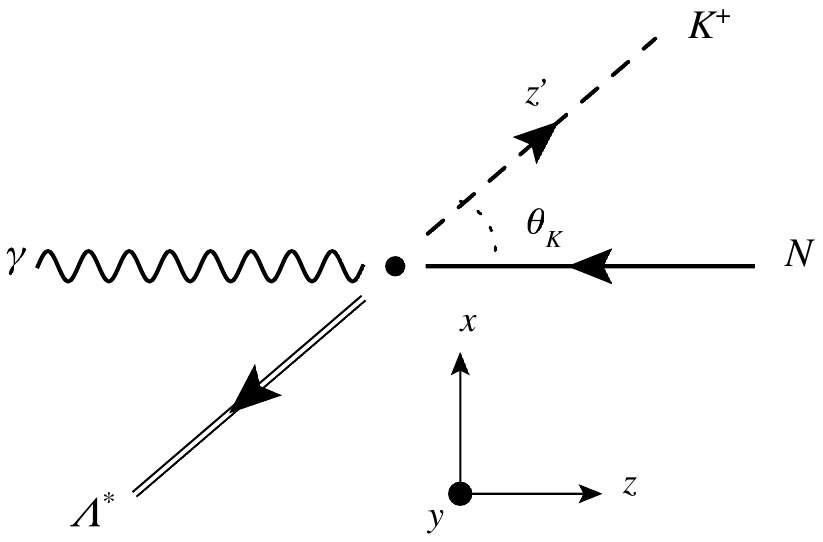}
\caption{Kinematical setup in the center of mass (cm) frame for the present work. The $z$- and $z'$-directions are taken along those of the incident photon and the outgoing kaon,  respectively. The angle between the photon and kaon is assigned as $\theta_{K}$.}       
\label{FIG0}
\end{figure}

Finally in this Section, we would like  to define the kinematical setup for computing the polarization transfer coefficients. The setup is schematically drawn in Fig.~\ref{FIG0}. We choose the $x$-$z$ plane as the reaction plane, and the $y$-axis is perpendicular to the plane. The $z$-axis is taken to be the direction of the incident photon in the center of mass (cm) system, whereas the angle between the photon and the outgoing kaon is assigned as the $\theta_{K}$. This kinematical setup is called the $(x,y,z)$-coordinate for convenience. On the contrary, if we choose the $z'$-axis along the outgoing kaon momentum, being equivalent to the rotation of the ($x,y,z$)-coordinate by $\theta_{K}$ in the clockwise direction, this is called the $(x',y',z')$-coordinate. The general rotation matrix for $(x,y,z,)\to(x',y',z')$ around  the $y$-axis can be written as
\begin{equation}
\label{eq:rot}
R_{y}=\left(
\begin{array}{ccc}
\cos\theta_{K}&0&-\sin\theta_{K}\\
0&1&0\\
\sin\theta_{K}&0&\cos\theta_{K}
\end{array}
 \right).
\end{equation}
\section{Brief review: unpolarized observables}
\begin{figure}[t]
\begin{tabular}{cc}
\includegraphics[width=7.0cm]{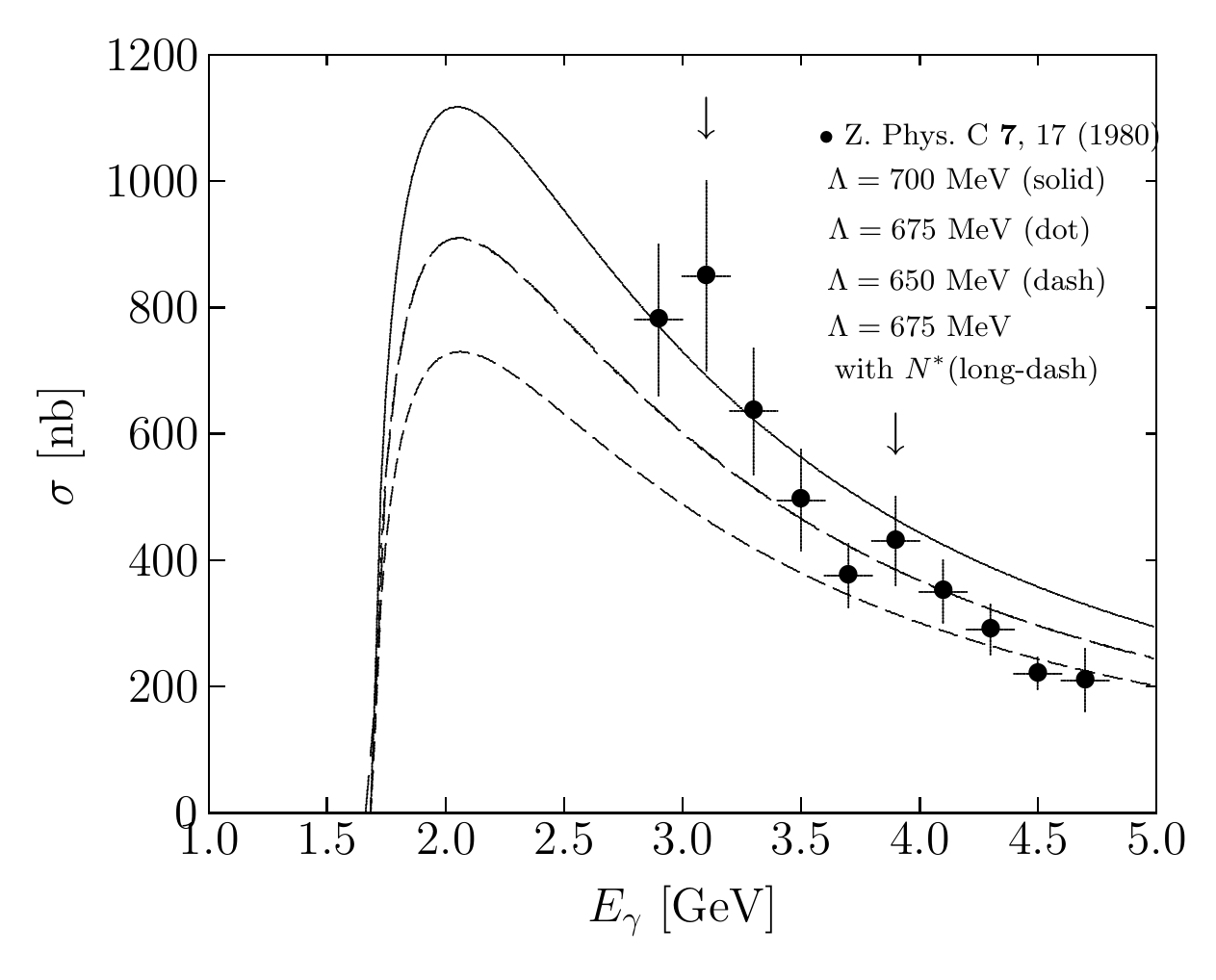}
\includegraphics[width=7.0cm]{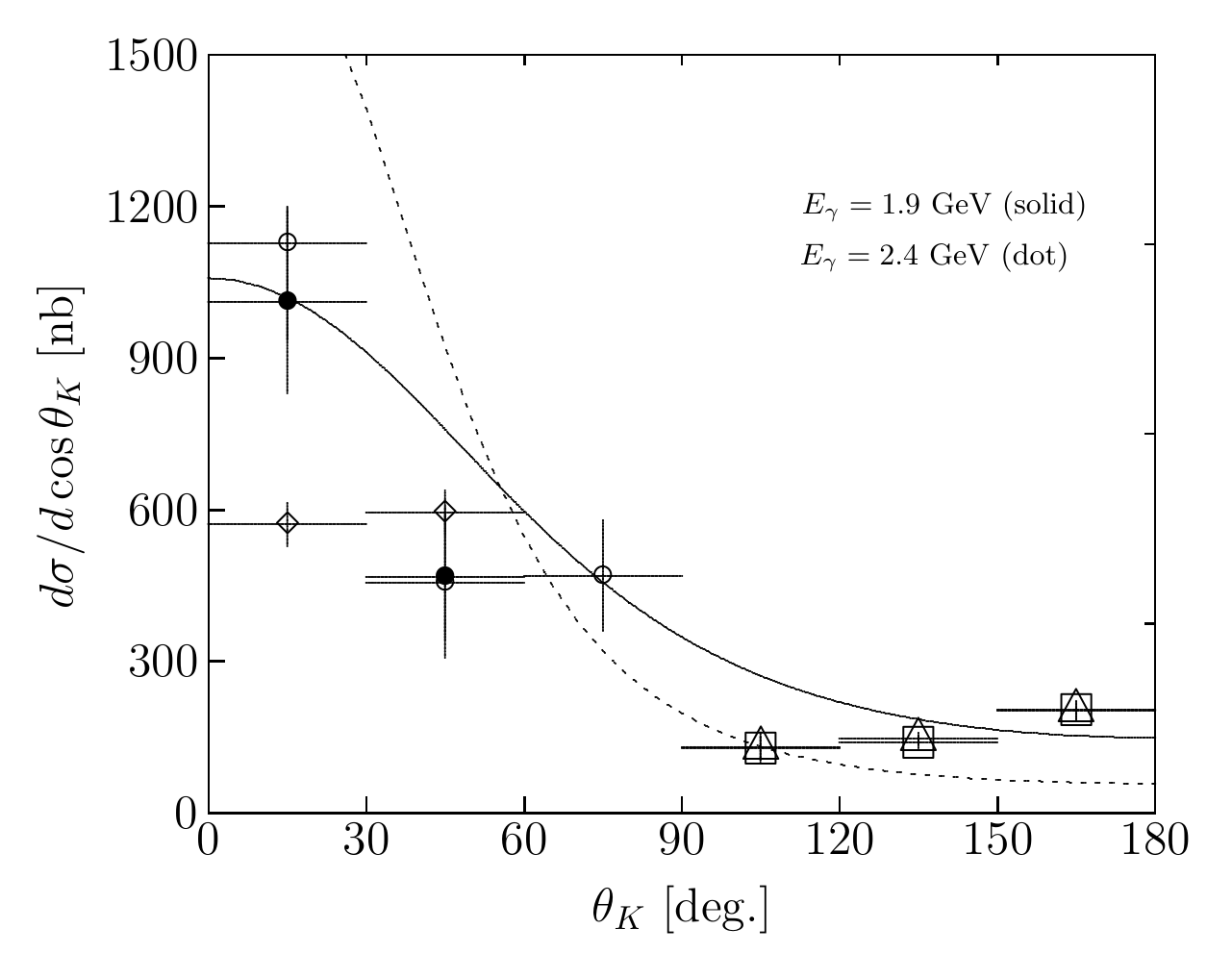}
\end{tabular}
\caption{Left: total cross section as a function of $E_{\gamma}$ for different cutoffs, $\Lambda=650$ MeV (dash), $675$ MeV (dot), and $700$ MeV (solid). We also draw the curve with $N^{*}(2080,3/2^{-})$ (long-dash) for $\Lambda=675$ MeV. The experimental data are taken from Ref.~\cite{Barber:1980zv}. Right: differential cross section as a function of $\theta_{K}$ using $\Lambda=675$ MeV at $E_{\gamma}=1.9$ GeV (solid) and $2.4$ GeV (dot) without the $N^{*}$ contribution. The experimental data are taken from Ref.~\cite{Muramatsu:2009zp}.}       
\label{UNPOL}
\end{figure}

Before computing the polarization transfer coefficients $C_{(x,z)}$, we want to make a brief review for the model calculations of the $\gamma N\to K\Lambda^{*}$ in the present frame work. Detailed calculations are given in our  previous work~\cite{Nam:2005uq}. In the left panel of Fig.~\ref{UNPOL}, we show the total cross section as a function of the photon energy $E_{\gamma}$. As for the solid, dotted, and dashed curves, we vary the cutoff parameter $\Lambda=650$, $675$, and $700$ MeV, respectively. Here, we only consider the Born terms without the $K^{*}$-exchange contribution. The experimental data comes from Ref.~\cite{Barber:1980zv}~\footnote{Recently, a new energy dependence for the $\Lambda^{*}$-photoproduction was reported in Ref.~\cite{Kohri:2009xe}, in which a possible narrow nucleon resonance was suggested to explain the data. However, in the present work, we do not consider such a new resonance contribution.}. 

All the cutoff values seem able to reproduce the data qualitatively well, depending on what energy region we are looking at. If we consider that the bumps indicated by the arrows at $E_{\gamma}\approx3.1$ GeV ($\sqrt{s}\approx2.6$ GeV) and $3.9$ GeV ($\sqrt{s}\approx2.9$ GeV) are coming from unknown resonance contributions, the curve with the $\Lambda=675$ MeV may be reasonable for the background from the Born terms. Thus, we will use this value for the cutoff hereafter for the numerical calculations. We also draw the total cross section with the resonance contribution from the $D_{13}$ with the long-dashed line, which is almost coincide with the dotted one. Here, the relative phase factor $\mathcal{C}$ in Eq.~(\ref{eq:re}) is chosen to be $+1$. From this result, we find that the $D_{13}$ contribution to the cross section in the present reaction process is almost negligible. We confirm that this behavior does not change much by altering the input parameters for the $D_{13}$. 

One of the most important consequences from our studies in Ref.~\cite{Nam:2005uq} is that the contact term dominates the process. Therefore, the $\Lambda^{*}$-photoproduction off the neutron target, in which the contact term does not appear, turns out to be much smaller than that off the proton one. This theoretical observation is the consequence of the gauge invariance, crossing symmetry, and on-shell condition for the form factors, although it may depend on the choice of the Lorentz structure for the spin-$3/2$ field. Interestingly, this finding has been supported experimentally~\cite{private} as well as theoretically~\cite{Toki:2007ab}. 

In the right panel of Fig.~\ref{UNPOL}, we depict the differential cross section as a function of the angle $\theta_{K}$. Here, we ignore the $K^{*}$-exchange contribution again. The experimental data are taken from Ref.~\cite{Muramatsu:2009zp}. The data were accumulated for the photon energy region $(1.9\lesssim E_{\gamma}\lesssim2.4)$ GeV. The data points indicated by the solid circle, open circle, diamond, triangle, and box are categorized by the different channels of the measured particles and analyzing methods. See Ref.~\cite{Muramatsu:2009zp} for more details. As shown in the figure, the theoretical calculations match well with the data qualitatively, although an obvious deviation can be found in the rise of the curve in the backward scattering region. We also find that the $D_{13}$ contribution is negligible for the differential cross section, although we do not show the resonance contribution in the figure. 

From these observations summarized in Fig.~\ref{UNPOL}, we can conclude that only the Born terms can reproduce the experimental data qualitatively very well and the resonance contribution may be negligible for the unpolarized observables, as long as we rely on the theoretical and experimental infformation for the resonances at this moment.     

\section{Polarization transfer coefficients}
In this Section, we define the polarization transfer coefficients $C_{(x,z)}$ for the $\Lambda^{*}$-photoproduction. Among the polarization observables in a meson photoproduction, the $C_{(x,z)}$ is identified as the spin asymmetry along the direction of the polarization of the recoil baryon with the circularly polarized photon beam. First, we define the polarization transfer coefficients in the $(x',y',z')$-coordiante, being similar to those for the spin-$1/2$ hyperon-photoproduction as in Refs.~\cite{McNabb:2003nf,Anisovich:2007bq}:
\begin{equation}
\label{eq:CXZ}
C_{x',|S_{x'}|}=
\frac{\frac{d\sigma}{d\Omega}_{r,0,+S_{x'}}
-\frac{d\sigma}{d\Omega}_{r,0,-S_{x'}}}
{\frac{d\sigma}{d\Omega}_{r,0,+S_{x'}}
+\frac{d\sigma}{d\Omega}_{r,0,-S_{x'}}},\,\,\,\,
C_{z',|S_{z'}|}=
\frac{\frac{d\sigma}{d\Omega}_{r,0,+S_{z'}}
-\frac{d\sigma}{d\Omega}_{r,0,-S_{z'}}}
{\frac{d\sigma}{d\Omega}_{r,0,+S_{z'}}
+\frac{d\sigma}{d\Omega}_{r,0,-S_{z'}}},
\end{equation}
where the subscripts $r$, $0$, and $\pm S_{x,'z'}$ stand for the right-handed photon polarization, unpolarized target nucleon, and polarization of the recoil baryon along the $x'$- or $z'$-axis, respectively. Since the photon helicity is fixed to be $+1$ here, the $C_{x',z'}$ measures the polarization transfer to the recoil baryon. Moreover, the $C_{x',z'}$ behave as the components of a three vector so that it can be rotated to the $(x,y,z)$-coordinate as:
\begin{equation}
\label{eq:ro}
\left(\begin{array}{c}
C_{x}\\C_{z}\end{array}\right)
=\left(
\begin{array}{cc}
\cos{\theta_{K}}&\sin{\theta_{K}}\\
-\sin{\theta_{K}}&\cos{\theta_{K}}
\end{array}
 \right)\left(\begin{array}{c}
C_{x'}\\C_{z'}\end{array}\right),
\end{equation}
where the $(x,y,z,)$-coordinate stands for that the incident photon momentum is aligned to the $z$-axis.

\begin{figure}[t]
\includegraphics[width=7.0cm]{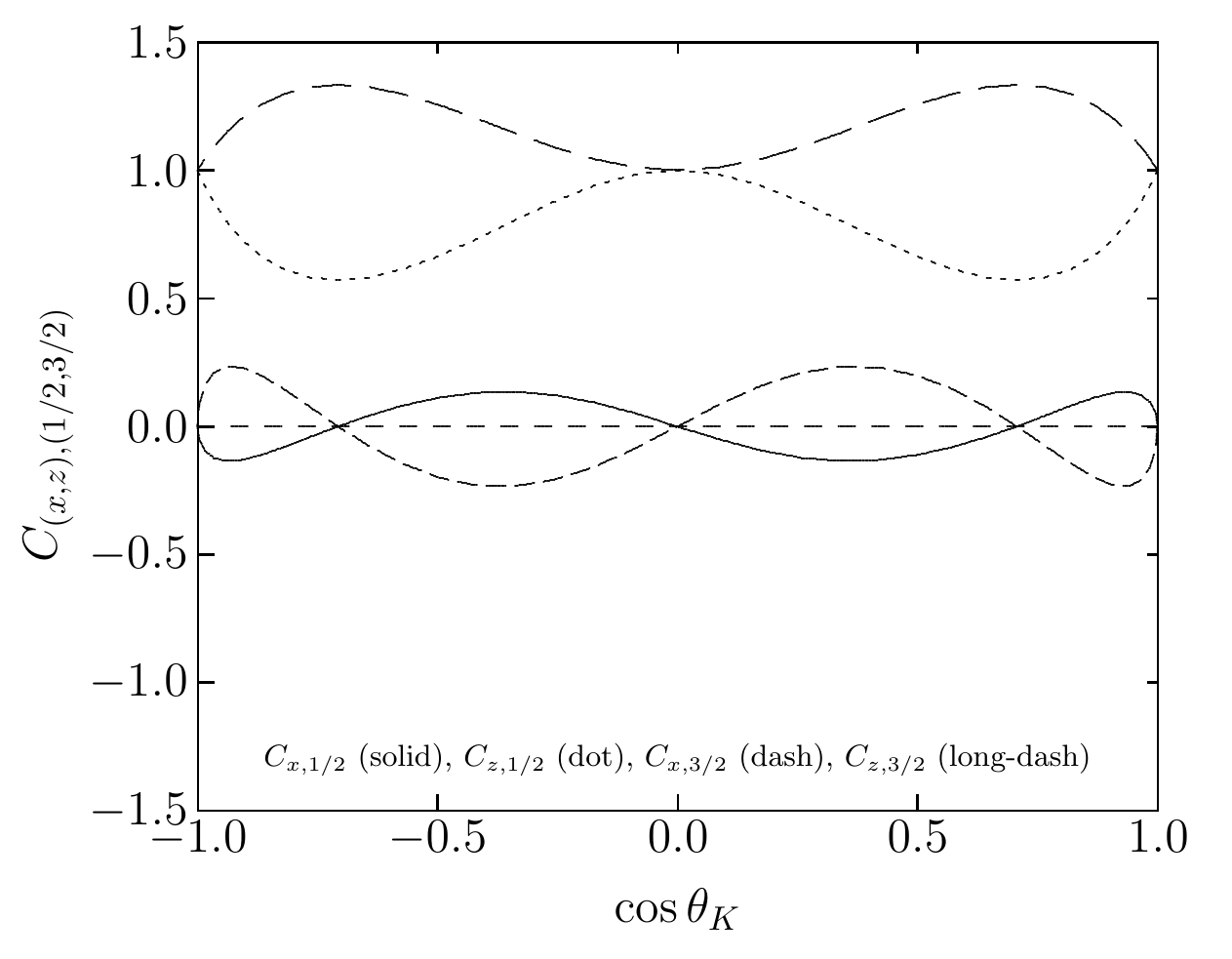}
\caption{$C_{(x,z),(1/2,3/2)}$ from the contact term only computation for the low-energy region.}
\label{FIG2}
\end{figure}

Here, we present a specific and simple example to compute the $C_{(x,z)}$ for the $\gamma N\to K\Lambda^{*}$ reaction process. As mentioned previously, since the contact term dominates the unpolarized amplitude and  gives a quite compact Lorentz structure as shown in Eq.~(\ref{eq:AMP}), it is a good starting point to consider only the contact term contribution for computing $C_{(x,z)}$ analytically. In the $(x',y',z')$-coordinate, when the polarization of the recoil baryon is directed to the $x'$- or $z'$-axis, it is easy to compute the $C_{x',z'}$ for the spin-$1/2$ state of the $\Lambda^{*}$, using the definition of the RS vector-spinor given in Appendix and the amplitude in Eq.~(\ref{eq:AMP}):
\begin{equation}
\label{eq:cxzprime}
C_{x',1/2}\sim-\frac{2s}{4c^{2}+s^{2}+1},\,\,\,\,
C_{z',1/2}\sim\frac{2c}{4s^{2}+c^{2}+1},
\end{equation}
where we have used abbreviations $c\equiv\cos{\theta_{K}}$ and $s\equiv\sin{\theta_{K}}$. Similarly, the spin-$3/2$ state of the $C_{(x',z')}$ can be obtained as
\begin{equation}
\label{eq:CP3}
C_{x',3/2}\sim-\frac{2s}{s^{2}+1},\,\,\,\,
C_{z',3/2}\sim\frac{2c}{c^{2}+1}.
\end{equation}
Note that these simple expressions for the $C_{(x',z')}$ are derived by assuming that the three momentum of the recoil baryon is much smaller than its mass (${\bm p}_{\Lambda^{*}}\ll M_{\Lambda^{*}}$). Using the rotation matrix Eq.~(\ref{eq:ro}), one can arrive at the following results for the $C_{(x,z)}$:
\begin{equation}
\label{eq:cxz1}
C_{x,1/2}\sim\frac{2sc}{4s^{2}+c^{2}+1}-\frac{2sc}{4c^{2}+s^{2}+1}
,\,\,\,\,
C_{z,1/2}\sim\frac{2c^{2}}{4s^{2}+c^{2}+1}+\frac{2s^{2}}{4c^{2}+s^{2}+1},
\end{equation}
and
\begin{equation}
\label{eq:cxz3}
C_{x,3/2}\sim\frac{2sc}{c^{2}+1}-\frac{2sc}{s^{2}+1},\,\,\,\,
C_{z,3/2}\sim\frac{2s^{2}}{s^{2}+1}+\frac{2c^{2}}{c^{2}+1}.
\end{equation}
It can be easily seen that the $C_{x}$ becomes zero, while the $C_{z}$ unity, in the collinear limit, $\theta_{K}=0^{\circ}$ or $180^{\circ}$, which corresponds to ${\bm p}_{\gamma}\parallel{\bm p}_{K}$. This is a natural and model-independent consequence of the helicity conservation of the scattering amplitude~\cite{Anisovich:2007bq}, i.e., no transverse polarization transfer at those angles. We depict the $C_{(x,z),(1/2,3/2)}$ in Eqs.~(\ref{eq:cxz1}) and (\ref{eq:cxz3}) in Fig.~\ref{FIG2} as functions of $\cos\theta_{K}$. Note that they are all symmetric around $\theta_{K}=90^{\circ}$. We also note that the curves  oscillate around their collinear values, zero or unity. 

\section{Numerical results}
\begin{figure}[t]
\begin{tabular}{cc}
\includegraphics[width=7.0cm]{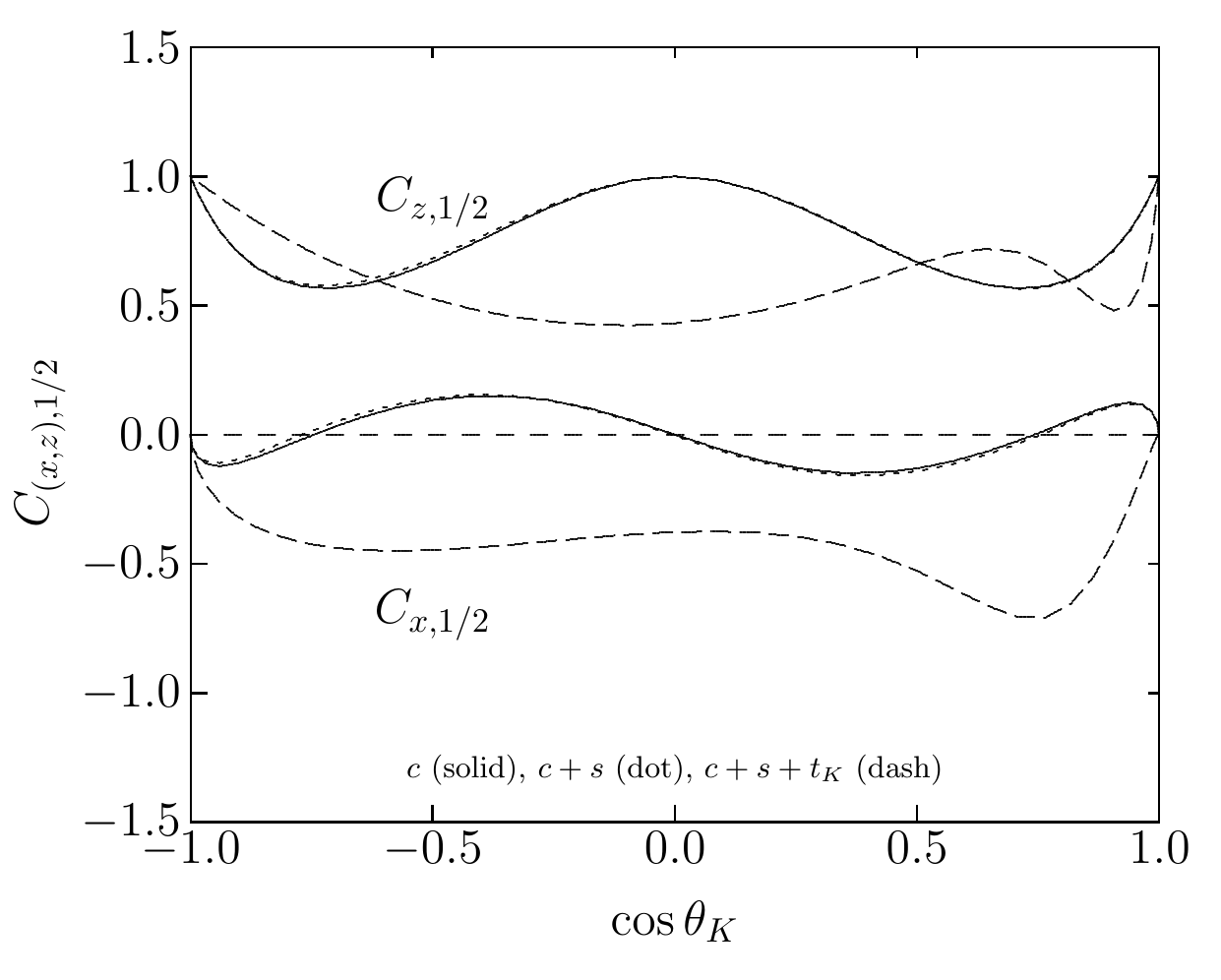}
\includegraphics[width=7.0cm]{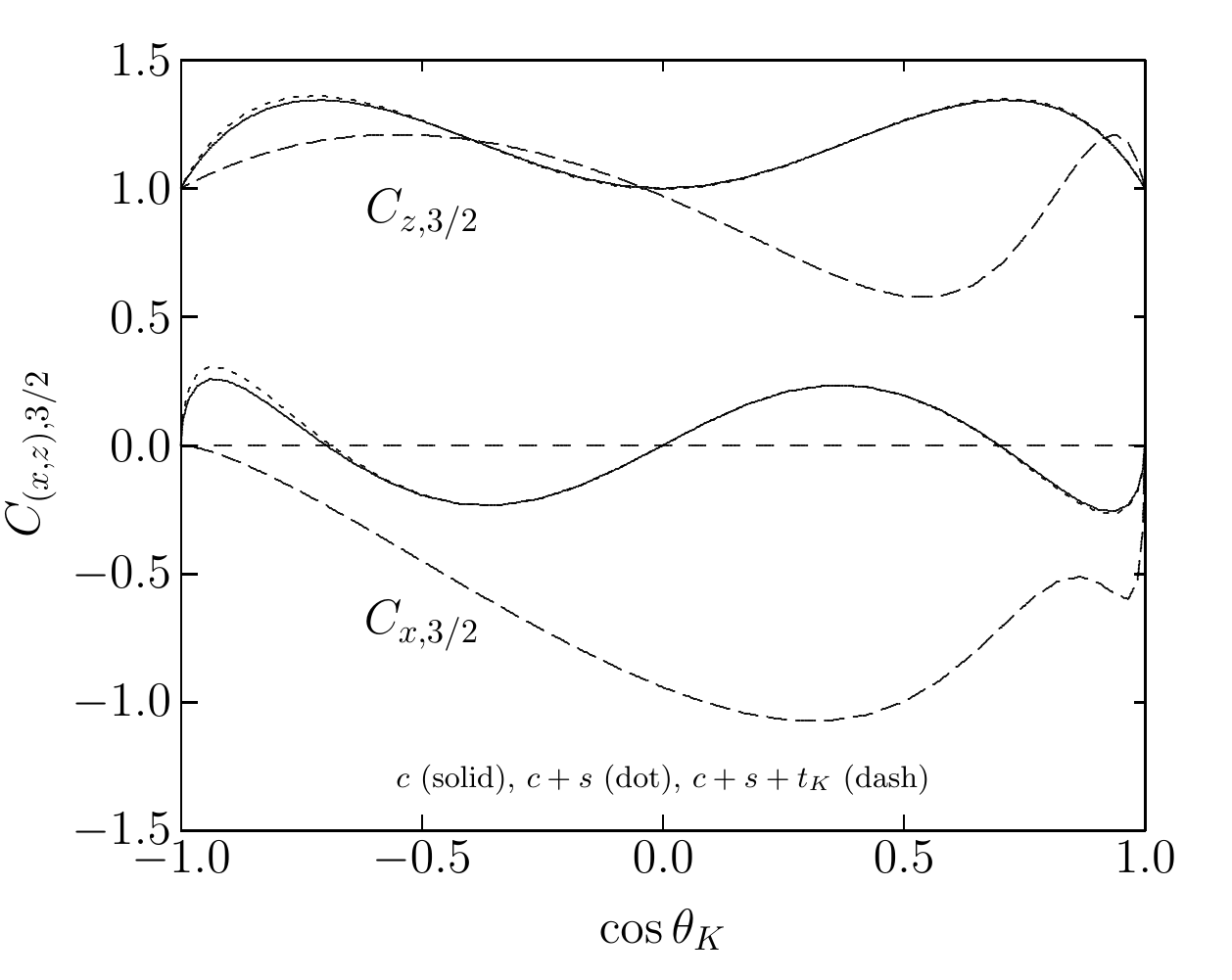}
\end{tabular}
\caption{Left: $C_{(x,y),1/2}$ as functions of $\cos\theta_{K}$ at $E_{\mathrm{cm}}=2.3$ GeV for each contribution: contact term (solid), contact term plus $s$-channel (dot), and adding the $K$-exchange (dash). Right: the same with the left panel but for $C_{(x,y),3/2}$. See text for details.}       
\label{FIG4}
\end{figure}

Now, we are in a position to provide numerical results for the polarization transfer coefficients $C_{(x,z),(1/2,3/2)}$ for the $\gamma p\to K^{+}\Lambda^{*}$ reaction process. First, we show the $C_{(x,z),(1/2,3/2)}$ curves at $E_{\mathrm{cm}}=2.3$ GeV for the separate contributions from each Born term for the spin-$1/2$ (left panel) and spin-$3/2$ (right panel) states in Fig.~\ref{FIG4}.
The solid lines stand for the case with the contact term only. As expected, they are almost the same with those given in Fig.~\ref{FIG2}. There are only negligible effects by adding the $s$-channel contribution (dotted line). The reason for this small effect can be understood by that the no angular dependence comes from the $s$-channel. As for the $u$-channel, we dropped it by taking its anomalous magnetic momentum $\kappa_{\Lambda^{*}}$ in Eq.~(\ref{eq:GROUND}) to be zero, since there is no available information on its value, and the $u$-channel is suppressed far more in comparison to other Born terms in the present form factor scheme~\cite{Nam:2005uq}. Interestingly, the inclusion of the $K$-exchange contribution changes the curves much for both the $C_{(x,z),1/2}$. In addition to the oscillation from the contact-term contribution, the interference between the contact and $K$-exchange contributions generates complicated angular dependence for the polarization transfer coefficients. This interesting behavior can be understood by that the $K$-exchange amplitude possesses much more complicated angular as well as energy dependent Lorentz structure in comparison to that for the contact term, as shown in Eq.~(\ref{eq:AMP}): $i\mathcal{M}^{K}_{t}\propto q_{t}\cdot e_{\Lambda^{*}}$, where the $q_{t}$ and $e_{\Lambda^{*}}$ represent the momentum transfer in the $t$-channel and $\Lambda^{*}$-polarization vector (see Appendix), respectively. According to the angular dependence from the $K$-exchange, we can see structures clearly in the vicinity around $\cos\theta_{K}\approx0.5$. We also draw the curves for the $C_{(x,z),3/2}$ in the right panel of the figure, and overall behaviors are similar to those for the $C_{(x,z),1/2}$.

\begin{figure}[t]
\begin{tabular}{cc}
\begin{tabular}{cc}
\includegraphics[width=7.0cm]{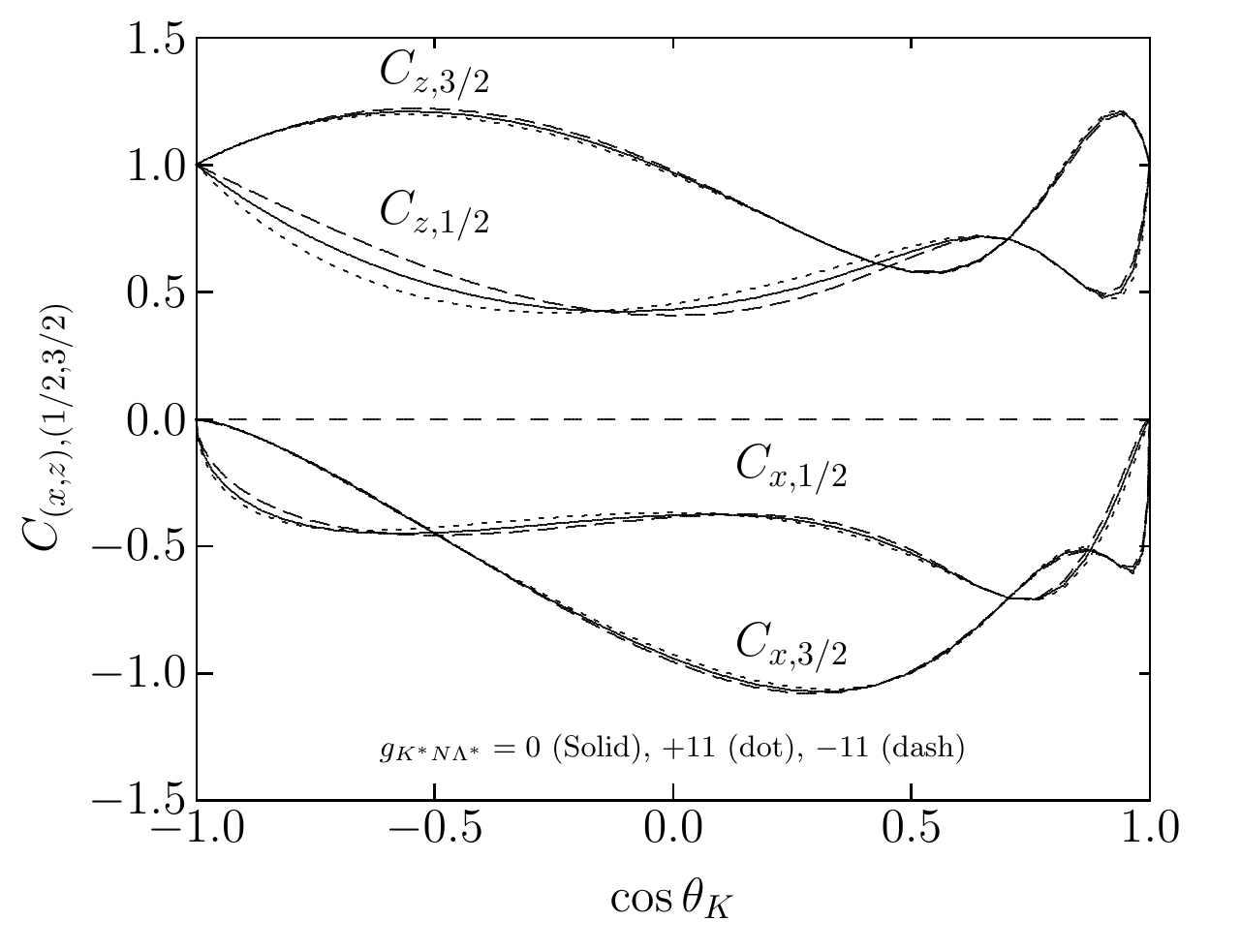}
\includegraphics[width=7.0cm]{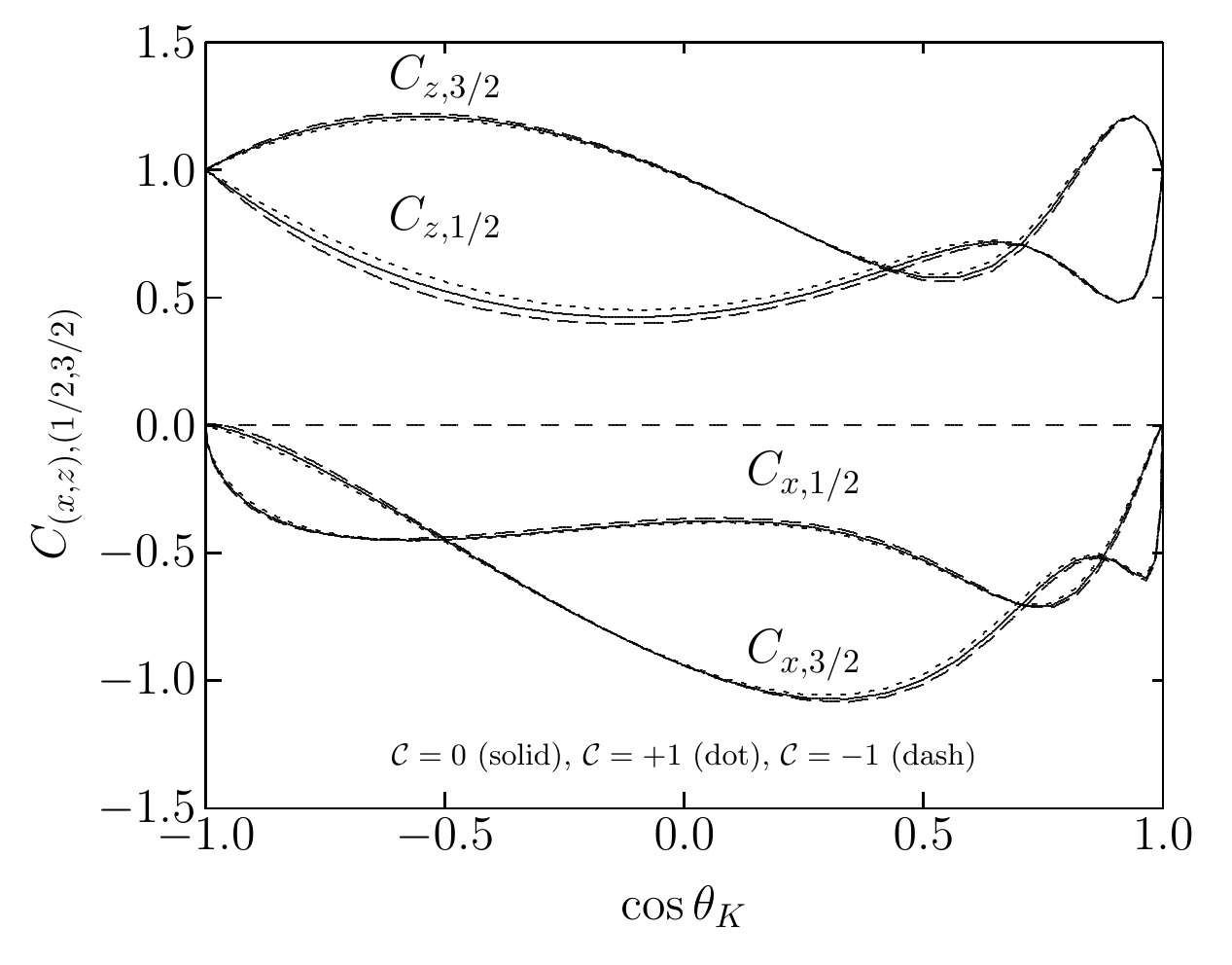}
\end{tabular}
\end{tabular}
\caption{Left: $C_{(x,y),(1/2,3/2)}$ as functions of $\cos\theta_{K}$ at $E_{\mathrm{cm}}=2.3$ GeV for $g_{K^{*}N\Lambda^{*}}=0$ (solid), $+11$ (dot), and $-11$ (dash). Right: $C_{(x,y),(1/2,3/2)}$ as functions of $\cos\theta_{K}$ at $E_{\mathrm{cm}}=2.3$ GeV for the $D_{13}(2080)$ resonance with $\mathcal{C}=0$ (solid), $\mathcal{C}=+1$ (dot), and $\mathcal{C}=-1$ (dash). See text for details.}       
\label{FIG41}
\end{figure}

In Fig.~\ref{FIG41}, we take into account the contributions from the $K^{*}$-exchange (left panel) and the $D_{13}$ resonance (right-panel). Among theoretical calculations, Ref.~\cite{Hyodo:2006uw} for instance, the strength of the $g_{K^{*}N\Lambda^{*}}$ was estimated much smaller than $g_{KN\Lambda^{*}}$. Hence, as a trial, we use $g_{K^{*}N\Lambda^{*}}=0,\pm11$. The numerical results for $E_{\mathrm{cm}}=2.3$ GeV reveal that the $K^{*}$-exchange contribution is very small in comparison to the contact and $K$-exchange as shown in the left panel. Only small deviation appears due to the $K^{*}$-exchange for the $C_{z,1/2}$ in the scattering region, $\cos\theta_{K}\lesssim0.5$. 

Similarly, it turns out that the effect from the $D_{13}(2080)$ is also negligible within our choice of the inputs for the resonance, where the $\mathcal{C}$ in the figure indicates the relative phase factor. Thus, we can ignore the $K^{*}$-exchange and $D_{13}$ contributions, in addition to the $s$- and $u$-channels, rather safely for further calculations. This irrelevance of such Born terms and resonance contribution leads us to an interesting result; the $C_{(x,z),(1/2,3/2)}$ in the present calculations are almost independent on the type of form factors and, in other words, represent the information of their bare amplitudes, since the form factor $F_{c}$ can be factorized and canceled from the $K$-exchange and contact term amplitudes as understood in Eqs.~(\ref{eq:AMP}) and (\ref{eq:CXZ}). 

\begin{figure}[t]
\begin{tabular}{cc}
\includegraphics[width=7.0cm]{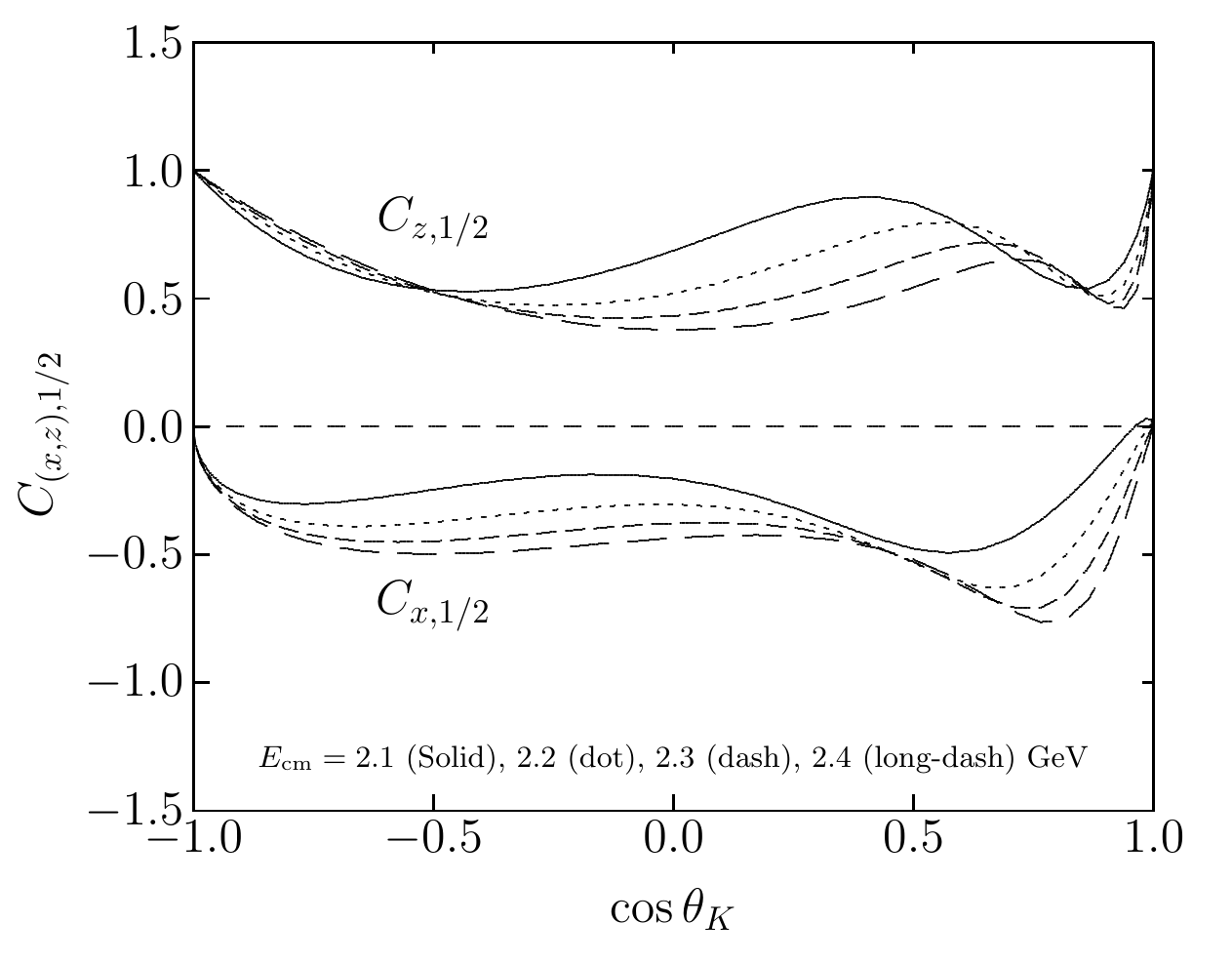}
\includegraphics[width=7.0cm]{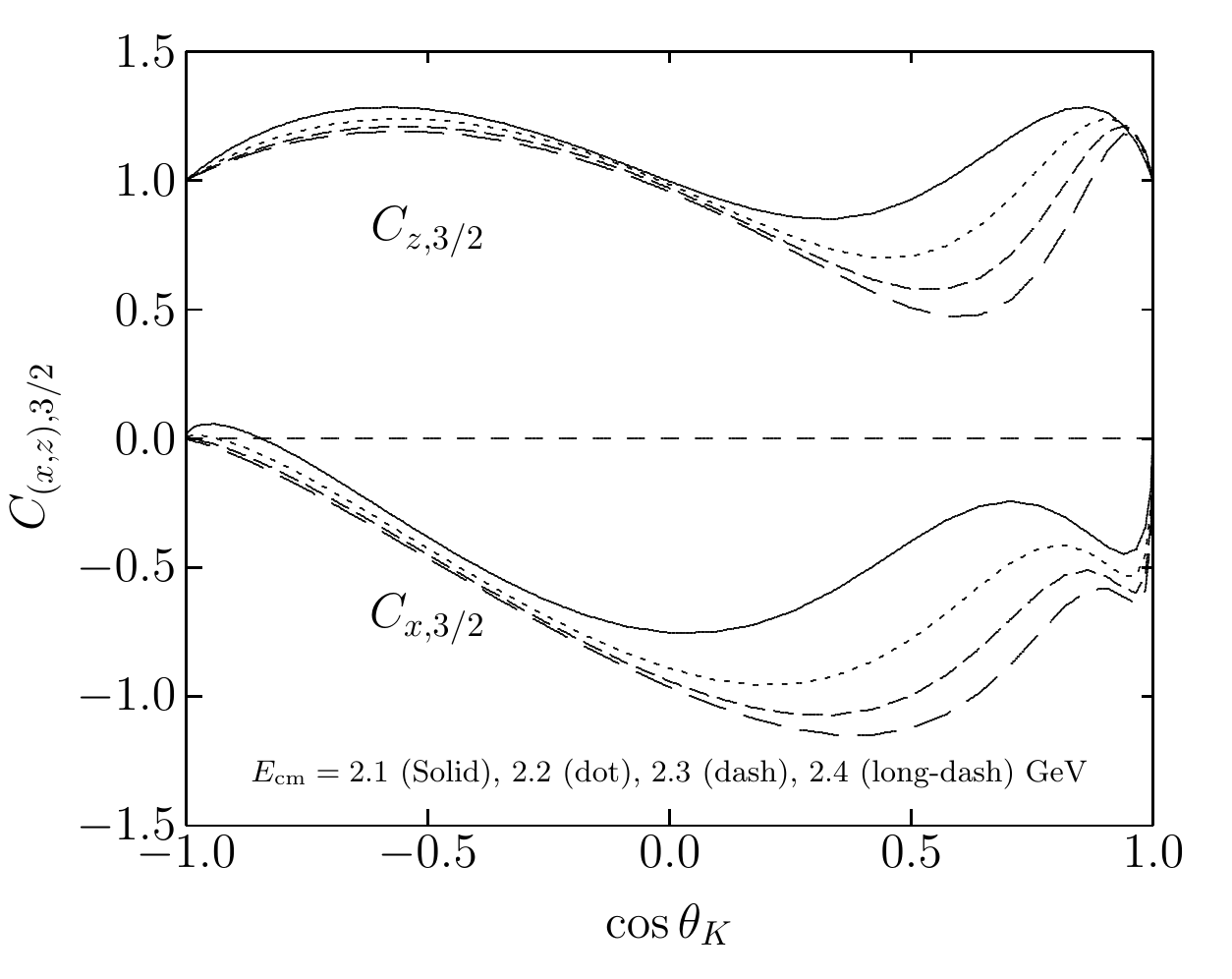}
\end{tabular}
\caption{$C_{(x,z),1/2}$ (left) and $C_{(x,z),3/2}$ (right) as functions of $\cos\theta_{K}$ without the $K^{*}$-exchange and $D_{13}(2080)$ contributions. The solid, dotted, dashed, long-dashed lines indicate the cases for $E_{\mathrm{cm}}=2.1$, $2.2$, $2.3$, and $2.4$ GeV, respectively.}       
\label{FIG5}
\end{figure}

From now on, we provide the full numerical results for the polarization transfer coefficients as functions of $\cos\theta_{K}$ as well as $E_{\mathrm{cm}}$. As noted above, we will not take into account the $K^{*}$-exchange and $D_{13}$ contributions. Since the present effective Lagrangian method is only applicable to the relatively low-energy regions, we do not go beyond $E_{\mathrm{cm}}\approx2.4$ GeV for the numerical calculations, since the reliability of the Born approximation can be broken down for the higher energy region. In Fig.~\ref{FIG5}, we show the numerical results for the $C_{(x,z),1/2}$ (left panel) and $C_{(x,z),3/2}$ (right panel) as functions of $\cos\theta_{K}$ for four different center of mass (cm) energies, $E_{\mathrm{cm}}=2.1$ (soild), $2.2$ (dot), $2.3$ (dash), and $2.4$ (long-dash) GeV. First. we take a look on the $C_{(x,z),1/2}$. Qualitatively, the angle dependence of the $C_{x}$ and $C_{z}$ are similar to each other. This tendency is very similar to the empirical observation $C_{z}-C_{x}\sim1$~\cite{Bradford:2006ba} for the ground state $\Lambda(1116)$-photoproduction. Moreover, as the energy increases, the polarization transfer to the recoil baryon $\Lambda^{*}$, $|C_{x,1/2}|$ also does, and vice versa for the $|C_{z,1/2}|$. Interestingly, for the angle region, $-0.5\lesssim\cos\theta_{K}\lesssim0.5$, the photon helicity is almost equally distributed to the $C_{x,1/2}$ and $C_{z,1/2}$. In the forward-scattering region ($\cos\theta_{K}\approx0$), these quantities are rapidly changed to zero or unity satisfying the collinear condition.  As for the $C_{(x,z),3/2}$ case, overall behaviors are similar to those of the $C_{(x,z),1/2}$ as shown in the right panel of Fig.~\ref{FIG5}. On the contrary, the change of the curves are much more obvious with respect to the angle as well as the energy. One can see that the polarization transfer to the $x$-axis is drastically increases for the region $0.25\lesssim\cos\theta_{K}\lesssim0.75$, according to the $K$-exchange contribution. 

\begin{figure}[t]
\begin{tabular}{cc}
\includegraphics[width=7.0cm]{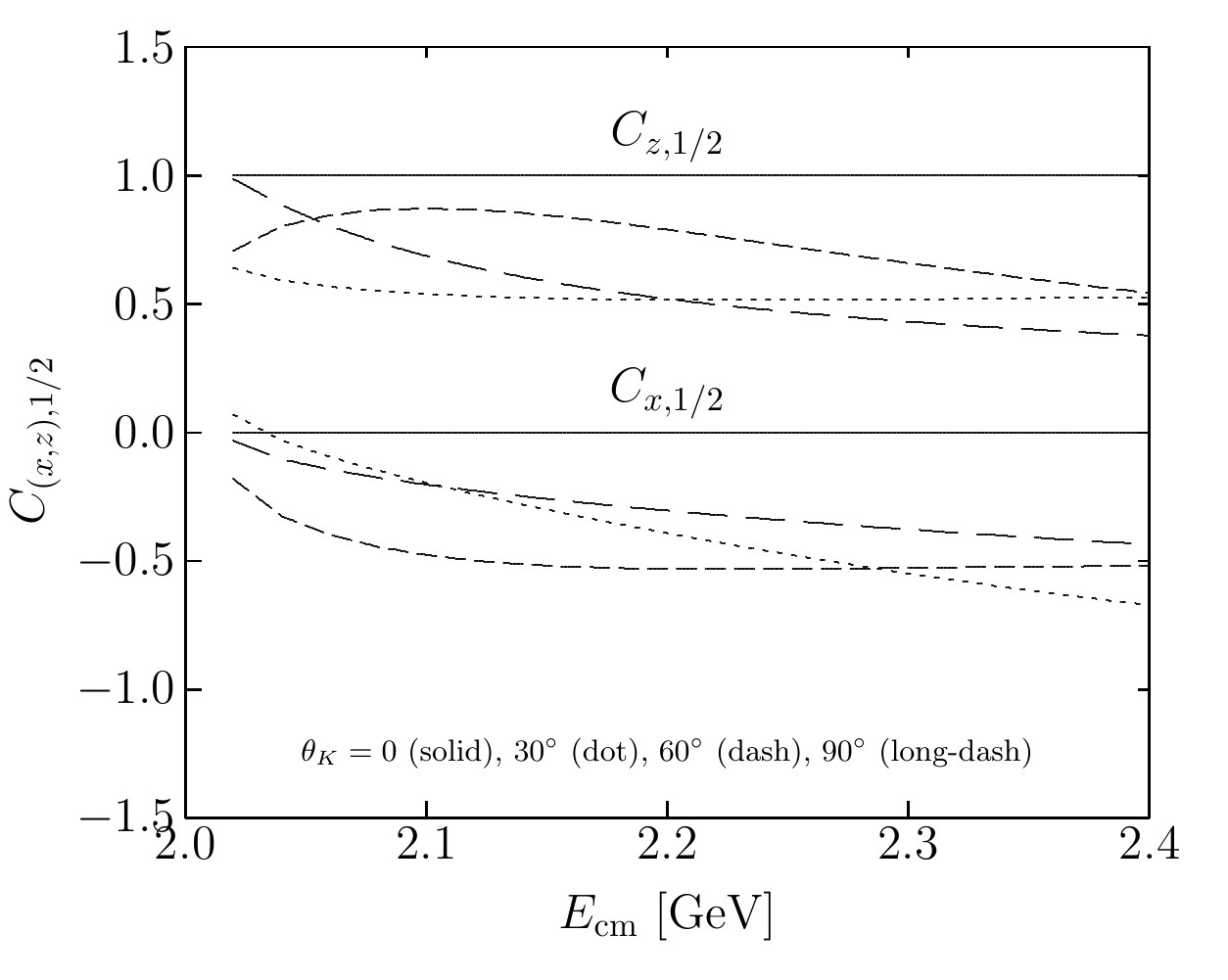}
\includegraphics[width=7.0cm]{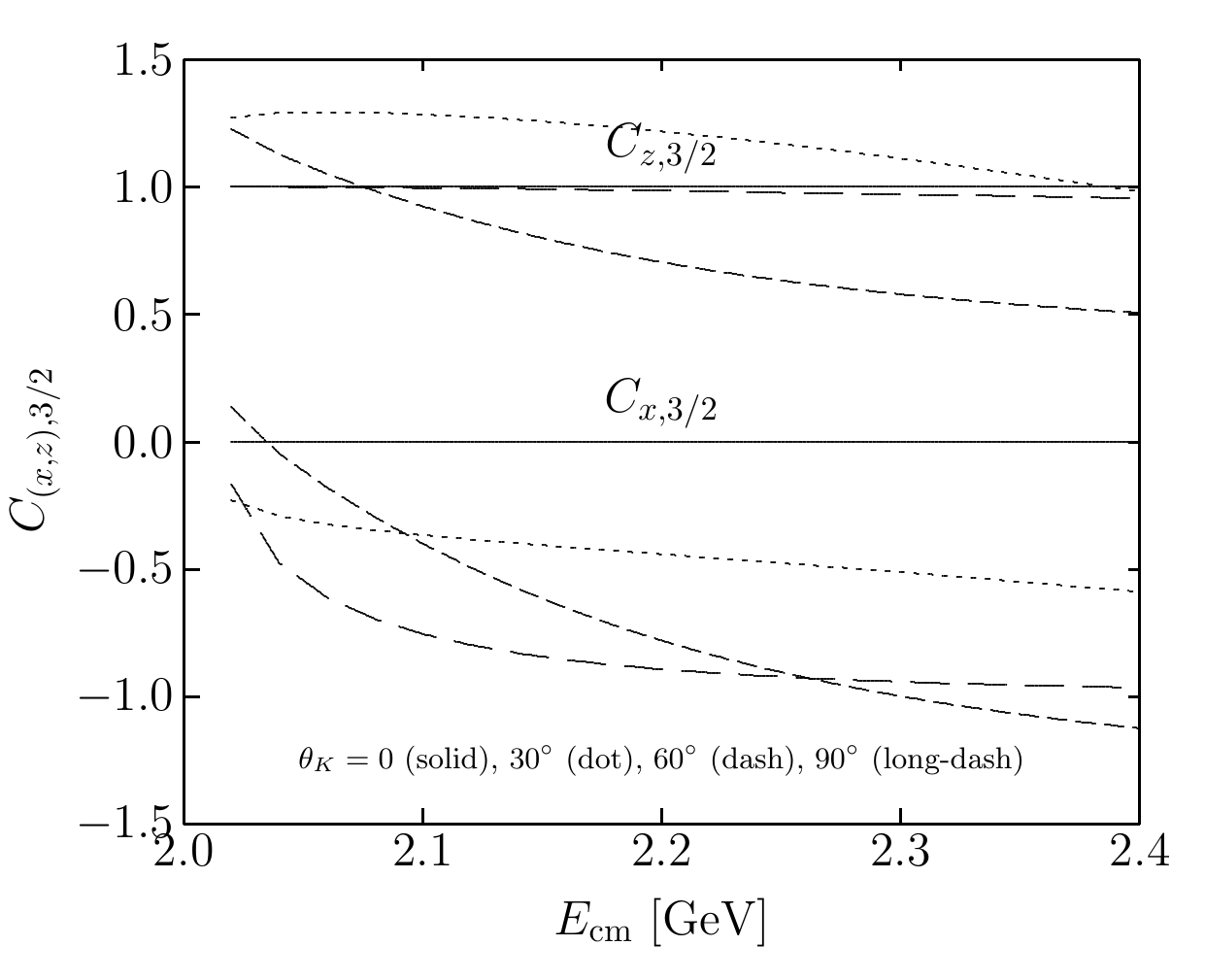}
\end{tabular}
\caption{$C_{(x,z),1/2}$ (left) and $C_{(x,z),3/2}$ (right) as functions of $E_{\mathrm{cm}}$ without the $K^{*}$-exchange and $D_{13}(2080)$ contributions. The solid, dotted, dashed, long-dashed lines indicate the cases for $\theta_{K}=0^{\circ}$, $30^{\circ}$, $60^{\circ}$, and $90^{\circ}$, respectively.}       
\label{FIG6}
\end{figure}

In Fig.~\ref{FIG6}, we show the numerical results for the $C_{(x,z),1/2}$ (left panel) and $C_{(x,z),3/2}$ (right panel) as functions of the cm energy for four different angles, $0^{\circ}$ (solid), $30^{\circ}$ (dot), $60^{\circ}$ (dash), and $90^{\circ}$ (long-dash) from $E_{\mathrm{cm}}=2.05$ GeV to $2.4$ GeV. As shown in the figure, at $\theta_{K}=0^{\circ}$, all the polarization transfers become zero or unity. As the angle increases, the complicated interference between the contact term and the $K$-exchange contributions takes place, resulting in that, as the energy increases, the $|C_{x}|$ also increases due to the $K$-exchange. Again, we can see the tendency $C_{z,1/2}-C_{x,1/2}\sim1$. 

\section{Summary and Conclusion}
In the present work, we have studied the polarization transfer coefficients $C_{(x,z)}$ for the $\vec{\gamma}N\to K^{+}\vec{\Lambda}(1520,3/2^{-})$ reaction process. The effective Lagrangian method was employed at tree level, and the gauge-invariant form factor scheme was taken into account. We also considered the nucleon resonance $D_{13}(2080)$ on top of the Born terms. We have observed the followings:
\begin{itemize}
\item All the numerical calculations show correct collinear behaviors: $C_{x,(1/2,3/2)}=0$ and $C_{z,(1/2,3/2)}=1$ at $\theta_{K}=0^{\circ}$ and $180^{\circ}$, respectively, due to the helicity conservation of the scattering amplitude. 
\item The $K^{*}$-exchange and $D_{13}(2080)$ contributions are very small as long as we employ the presently available experimental and theoretical information on them.
\item The contact-term contribution plays a significant role to determine the $C_{(x,z)}$ and produce their basic shapes, which oscillate around the collinear values, zero and unity, as functions of $\cos\theta_{K}$. 
\item The $K$-exchange contribution enhances the transverse polarizaztion transfer to the $\Lambda^{*}$ along the $x$-axis, according to the interference with the contact-term, as the $E_{\mathrm{cm}}$ increases. 
\item The $C_{(x,z),(1/2,3/2)}$ show structures around $\theta_{K}\approx60^{\circ}$ due to the $K$-exchange contribution, and $|C_{x,(1/2,3/2)}|$ and $|C_{z,(1/2,3/2)}|$ increase and decrease with respect to $E_{\mathrm{cm}}$. 
\item This contact term and $K$-exchange dominance reduces the theoretical uncertainty from the form factors in the present approach.
\end{itemize}
Since the CLAS and LESP collaborations have already explored the $\Lambda(1520)$-photoproduction for the unpolarized observables and are possible to produce the linearly or circularly polarized photon beam, we expect that polarization observables such as the $C_{(x,z),(1/2,3/2)}$ will be measured in near future. Then, the present theoretical calculations will be a useful guidance for analyzing the experimental data. At the same time, by comparing with the data, we can test the present theoretical approach for the $\Lambda^{*}$-photoproduciton more precisely. 

\section*{Acknowledgment}
The author thanks W.~C.~Chang, A.~Hosaka, H.~-Ch.~Kim, T.~Nakano, C.~W.~Kao, and H.~Kohri for fruitful discussions. This work was supported by the NSC96-2112-M033-003-MY3 from the National Science Council (NSC) of Taiwan. The numerical calculations were performed using MIHO at RCNP, Osaka University, Japan.
\section*{Appendix}
The Rarita-Schwinger vector-spinor with the four momentum $p=(E,{\bm p})$, satisfying $p^{2}=M^{2}$, reads
\begin{eqnarray}
\label{eq:rs}
u^{\mu}(p,+\frac{3}{2})&=&e^{\mu}_{+}(p)\,u(p,+\frac{1}{2}),
\cr
u^{\mu}(p,+\frac{1}{2})&=&
\sqrt{\frac{2}{3}}e^{\mu}_{0}(p)\,u(p,+\frac{1}{2})
+\sqrt{\frac{1}{3}}e^{\mu}_{+}(p)\,u(p,-\frac{1}{2}),
\cr
u^{\mu}(p,-\frac{1}{2})&=&\sqrt{\frac{1}{3}}e^{\mu}_{-}(p)\,u(p,+\frac{1}{2})
+\sqrt{\frac{2}{3}}e^{\mu}_{0}(p)\,u(p,-\frac{1}{2}),
\cr
u^{\mu}(p_{2},-\frac{3}{2})&=&e^{\mu}_{-}(p)\,u(p,-\frac{1}{2}),
\end{eqnarray}
where the polarization four-vector with the helicity $\lambda$ defined by
\begin{equation}
\label{eq:polvec}
e^{\mu}_{\lambda}(p)=
\left(\frac{{\bm e}_{\lambda}\cdot{\bm p}}{M},\,\,\,\,\,
{\bm e}_{\lambda}
+\frac{{\bm e}_{\lambda}\cdot{\bm p}}{M}
\frac{{\bm p}}{E+M}\right).
\end{equation}
Here the polarization three-vector is written as
\begin{equation}
\label{eq:polar}
{\bm e}_{+}=-\frac{1}{\sqrt{2}}(\cos\theta,i,-\sin\theta),
\,\,\,\,
{\bm e}_{0}=(\sin\theta,0,\cos\theta)
\,\,\,\,
{\bm e}_{-}=\frac{1}{\sqrt{2}}(\cos\theta,-i,-\sin\theta).
\end{equation}
These unit vectors indicate that the polarization is pointing to the ${\bm e}_{0}$-direction. Then, the four-component spinors along this direction are defined as:
\begin{equation}
\label{eq:spinor}
u(p,+\frac{1}{2})=
\sqrt{E+M}\left(
\begin{array}{c}
\cos\frac{\theta}{2}\\
\sin\frac{\theta}{2}\\
\frac{p_{3}\cos\frac{\theta}{2}+(p_{1}-ip_{2})\sin\frac{\theta}{2}}{E+M}\\
\frac{-p_{3}\sin\frac{\theta}{2}+(p_{1}+ip_{2})\cos\frac{\theta}{2}}{E+M}
\end{array} \right),\,\,\,\,
u(p,-\frac{1}{2})=
\sqrt{E+M}\left(
\begin{array}{c}
\cos\frac{\bar{\theta}}{2}\\
\sin\frac{\bar{\theta}}{2}\\
\frac{p_{3}\cos\frac{\bar{\theta}}{2}+(p_{1}-ip_{2})
\sin\frac{\bar{\theta}}{2}}{E+M}\\
\frac{-p_{3}\sin\frac{\bar{\theta}}{2}+(p_{1}+ip_{2})
\cos\frac{\bar{\theta}}{2}}{E+M}
\end{array} \right),
\end{equation}
where $\bar{\theta}=\theta-\pi$. We take $\theta=\theta_{K}$ and $\theta_{K}+\frac{\pi}{2}$ for the recoil-baryon polarizations along $z'$- and $x'$-axes, respectively. 


\end{document}